\documentclass[aps,a4paper,twocolumn,pra]{revtex4-1}
\pdfoutput=1

\usepackage{amsmath}%
\usepackage{amsfonts}%
\usepackage{amssymb}%
\usepackage[cal=cm]{mathalfa}
\usepackage{graphicx}
\usepackage{physics}
\usepackage{sidecap}
\usepackage{color}
\usepackage{bbm}
\usepackage{nicefrac}
\usepackage{float}
\usepackage[dvipsnames]{xcolor}
\usepackage[utf8]{inputenc}
\usepackage[normalem]{ulem}
\usepackage{nicefrac}
\newcommand{\stkout}[1]{\ifmmode\text{\sout{\ensuremath{#1}}}\else\sout{#1}\fi}

\usepackage[citebordercolor=blue]{hyperref}

\begin{document}

\title{Additive Manufactured and Topology Optimized Permanent Magnet Spin-Rotator for Neutron Interferometry Applications}

\author{Wenzel Kersten$^1$}
\author{Laurids Brandl$^1$}
\author{Richard Wagner$^1$}
\author{Christian Huber$^{2,3}$}
\author{Florian Bruckner$^{2,3}$}
\author{Yuji Hasegawa$^{1,4}$}
\author{Dieter Suess$^{2,3}$}
\author{Stephan Sponar$^1$}


\affiliation{%
$^1$Atominstitut, TU Wien, Stadionallee 2, 1020 Vienna, Austria \\
$^2$Physics of Functional Materials, University of Vienna, 1090 Vienna, Austria \\
$^3$Christian Doppler Laboratory for Advanced Magnetic Sensing and Materials, 1090 Vienna, Austria \\
$^4$Department of Applied Physics, Hokkaido University, Kita-ku, Sapporo 060-8628, Japan}

\date{\today}

\hyphenpenalty=800\relax
\exhyphenpenalty=800\relax
\sloppy
\setlength{\parindent}{0pt}

\noindent

\begin{abstract}
In neutron interferometric experiments using polarized neutrons coherent spin-rotation control is required. In this letter we present a new method for Larmor spin-rotation around an axis parallel to the outer guide field using topology optimized 3D printed magnets. The use of 3D printed magnets instead of magnetic coils avoids unwanted inductances and offers the advantage of no heat dissipation, which prevents potential loss in interferometric contrast due to temperature gradients in the interferometer. We use topology optimization to arrive at a design of the magnet geometry that is optimized for homogeneity of the magnetic action over the neutron beam profile and adjustability by varying the distance between the 3D printed magnets. We verify the performance in polarimetric and interferometric neutron experiments.
\end{abstract}

\maketitle

\begin{SCfigure*}[]
	\centering
	\includegraphics[width=0.75\textwidth]{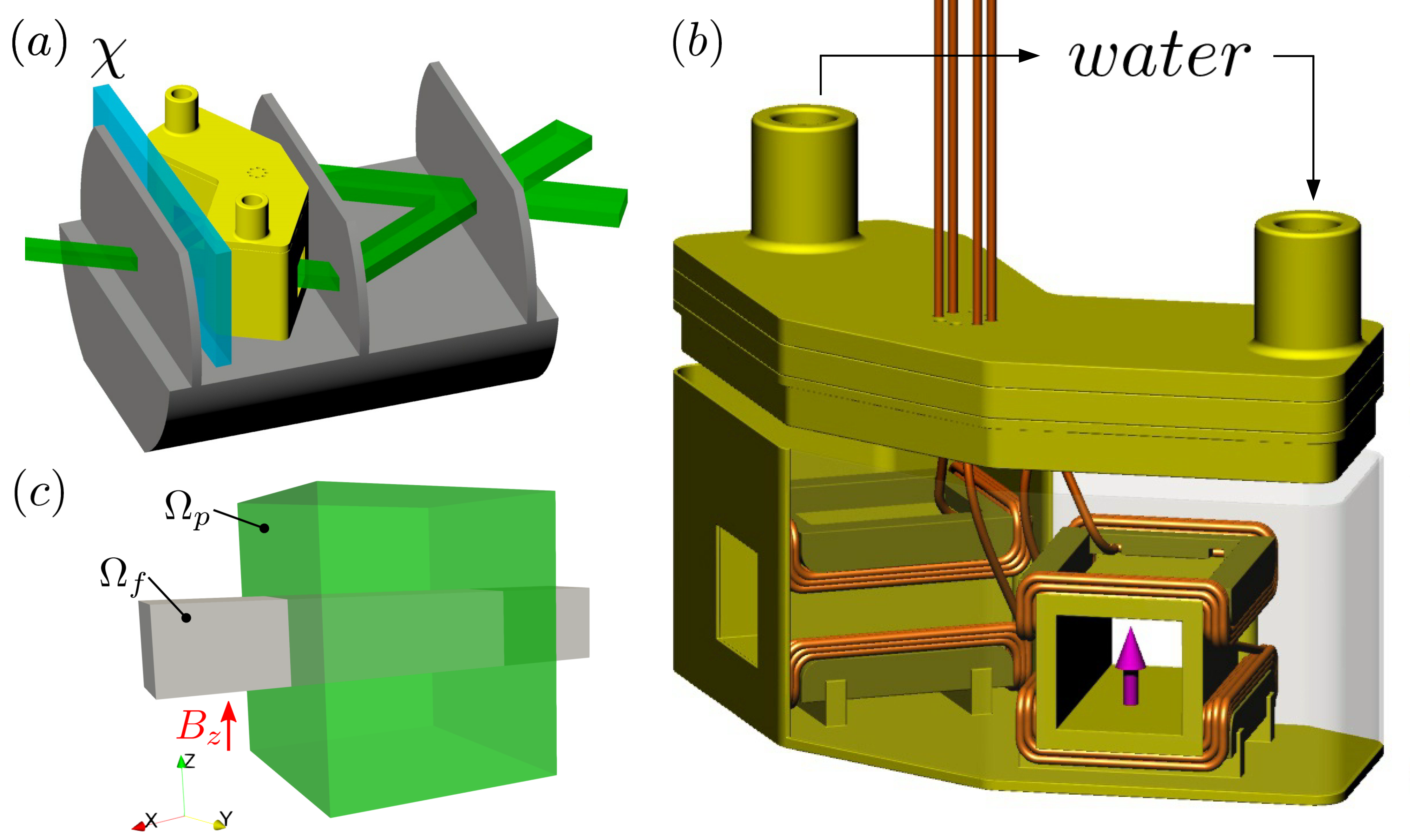}
	\caption[Current neutron interferometric setup.]{Current neutron interferometric setup. (a) Placement of the box inside the neutron interferometer together with the phase shifter plate $\chi$. 
	(b) Schematic view of the water-cooled Larmor spin-rotator box with coils in Helmholtz geometry and magnetic field direction indicated in magenta.  Also depicted are in and outlets for temperature-controlled cooling water. The box needs to be waterproof, which makes manufacture tedious. (c) Design and field domain for an optimized 3D printed permanent magnetic Larmor spin-rotator.}
	\label{fig:interferometer_2}
\end{SCfigure*}

\section{Introduction}
Since its first demonstration in 1974, neutron interferometry \cite{Rauch74,RauchBook} has proven itself a powerful tool to study fundamental concepts of quantum mechanics. For instance the spin-superposition law \cite{Badurek83Direct,Badurek83TimeDepend}, gravitationally induced phases (Colella-Overhause-Werner effect) \cite{Colella75} or the Sagnac effect \cite{Werner79} have been verified experimentally.
Experiments challenging our views on reality include the Einstein-Podolsky-Rosen experiments, where a violation of Bell's inequality, assuming local hidden variables, can be measured. Using pairs of entangled photons \cite{BookBertlmannZeilinger,Freedman72,Aspect82,Kwiat95,Weihs98,Tittel98,Rowe01,Zeilinger15,Hensen15} it can be shown that quantum mechanics is non-local, i.e., it cannot be reproduced by local realistic theories. Entanglement can not only be achieved for spatially separated particles but also between different degrees of freedom \cite{Moehring04,Matsukevich2008,hasegawa2003violation,Sakai06,Geppert18}, e.g. path, spin and energy in the case of neutrons. This enables the experimental violation of Bell's inequality for non-contextual hidden variables using neutron interferometric setups \cite{Hasegawa06,Bartosik09}, giving proof that measurement outcomes are not predetermined and therefore depend on the experimental context \cite{Bell66,Mermin93}. Not a statistical violation but a contradiction between quantum mechanics and local hidden variable theories was found by Greenberger, Horne and Zeilinger in 1989 for, at least, tripartite entanglement \cite{GHZ89Pro,GHZ90}, also feasible with neutrons \cite{Hasegawa10}.
More recently, a neutron optical approach for obtaining weak values \cite{Sponar15}, a new type of quantum variable introduced by Aharanov already three decades ago \cite{Aharonov88}, has been realized. With this novel technique quantum paradoxes such as the Quantum Cheshire Cat \cite{Denkmayr14} or a violation of the classical Pigeonhole effect \cite{Cai17} have been studied. In addition, a direct tomographic state reconstruction technique based on weak values, independent of the measurement strength, has been established \cite{Denkmayr17}. Common to all these experiments is the requirement of high interference contrast and even more important, a high efficiency in spin manipulation capacity, i.e., full control of the spin state.

\section{3D printed spin-rotator}

The current setup to tests of fundamental phenomena in quantum mechanics
 Bell's inequality violation at the Institut Laue Langevin in Grenoble, France, is shown in Fig.\,\ref{fig:interferometer_2}. The Larmor spin-rotator has a  Helmholtz coil geometry.  It applies a local field in addition to the guide field in the $z$-direction, thereby locally changing the Larmor frequency, with which the spin precesses around the field in the $xy$-plane. The rotation angle is given by
\begin{align}
 \alpha(B_z) = \frac{2\mu_N}{\hbar} B_z  \frac{l}{v}
 \label{eq:alpha}
\end{align}
where $\mu_N=-9.6623647\times10^{-27}$~J/T is the magnetic moment of a neutron, $l$ is the length of the coil and $v$ is the velocity of the neutrons. Since the local field and the guide field are parallel, the field transition can be adiabatic.

Such a Larmor spin-rotator must have a highly homogeneous field along the neutron beam path, because an inhomogeneous field would lead to a dephasing of the neutron and therefore to a loss in contrast of the interferogram. Another crucial point is the thermal stability of the setup. A change of temperature during the measurement leads to a loss in contrast, as phase drifts occur, e.g., a temperature change of $1$~$^\circ$C results in $1.92$~rad phase shift \cite{geppert2014improvement}. For this reason the Helmholtz coil Larmor spin-rotator is water-cooled (Fig.\,\ref{fig:interferometer_2}(b)), which complicates the setup of the experiment, because the temperature of the cooling water has to be optimized. In addition, the manufacture of waterproof boxes to hold the Helmholtz coils is tedious.

\subsection{Requirements}
\begin{figure}[!b]
	\centering
	\includegraphics[width=0.45\textwidth]{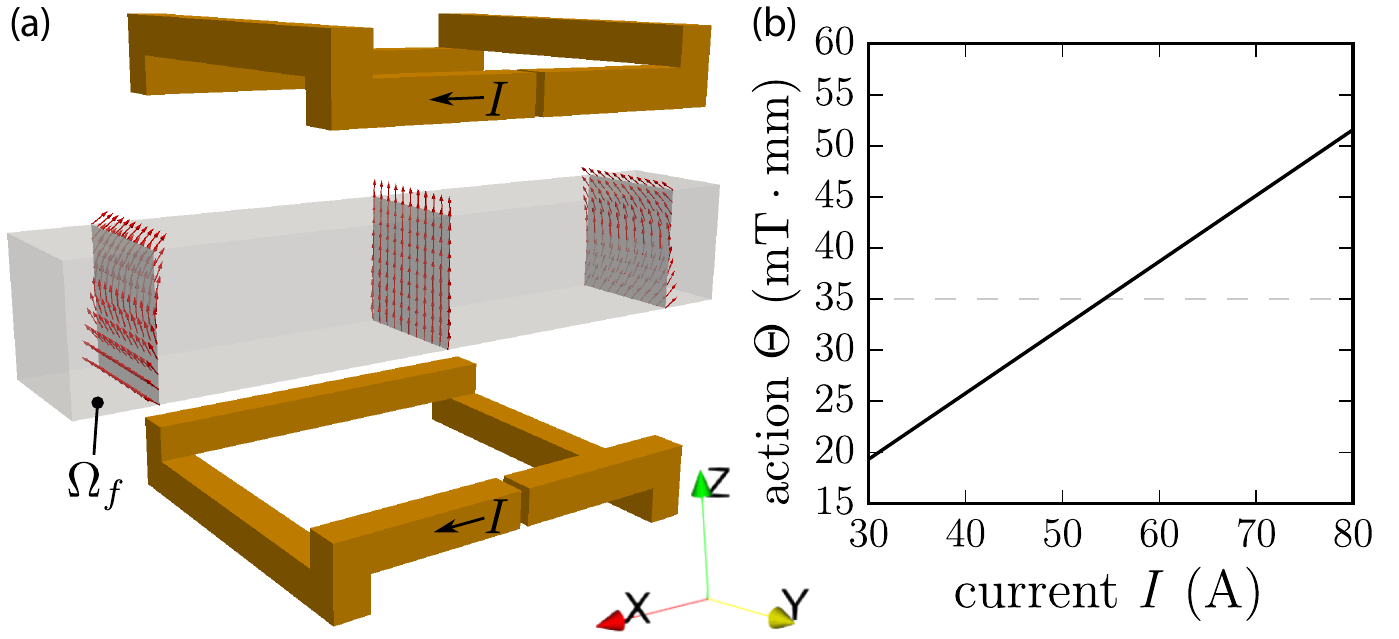}
	\caption[Current driven Larmor spin-rotator.]{Current driven Larmor spin-rotator. (a) Modified Helmholtz coil geometry with the field box $\Omega_f$ (equal to neutron beam size). Normalized magnetic field vectors in red. (b) Action $\Theta$ as a function of the current $I$ through the Helmholtz coil.}
	\label{fig:spinflipper_helmholz}
\end{figure}

To improve the design at hand, the actual condition should be examined in more detail. The neutron path is simplified by a field box $\Omega_f$ with a size of $7\times7\times40$~mm$^3$ ($a\times a\times L$). To describe the influence of the magnetic field on the phase shift of the neutrons, the action $\Theta$ is defined as 
\begin{align}
\Theta=\frac{1}{a^2}\int_{\Omega_{f}}| B_z |  \mathrm{d} \boldsymbol{r}.
\end{align}
In order to rotate the neutron spin by an angle $\alpha=\pi$, an action of $\Theta=35~\text{mT}\cdot\text{mm}$ is necessary. The action of the Helmholtz coil geometry is calculated using the finite element method (FEM) tool Magnum.fe \cite{magnumfe}. Fig.~\ref{fig:spinflipper_helmholz}(a) shows the geometry as well as the field box with vectors of the magnetic field. The field shows an inhomogeneous behavior outside the coil. A calculation of the current $I$ to reach an action of $\Theta=35~\text{mT}\cdot\text{mm}$ is plotted in Fig.\,\ref{fig:spinflipper_helmholz}(b). The relative error of the field homogeneity is plotted in \ref{fig:gap_mag_action}(b). An optimized design should be found with the requirements, presented in the next subsection. 

The goal is to find an optimized permanent magnetic Larmor spin-rotator by the usage of the inverse stray field and topology optimization framework \cite{bruckner2017solving,huber20173d,huber2017topology}. Following design parameters are given:
  	 \begin{itemize}
	 \item Size of the field box $\Omega_f$: $7\times7\times40$~mm$^3$ ($a\times a\times L$).
	 \item Maximum design volume $\Omega_p$: $24\times24\times20$~mm$^3$ ($a\times a\times L$).
	 \item Action $\Theta$ of the external field: $\Theta=35$~mT$\cdot$mm.
	 \item $\Theta$ is adjustable in the range of $\pm 5$~mT$\cdot$mm.
	 \item Homogeneous magnetic field density $\boldsymbol{B}(\boldsymbol{r})$ along $z$-axis: $\boldsymbol{B}(\boldsymbol{r})=(0,0,B_z)$.
	\end{itemize}

A challenge for a permanent magnetic system is to make the action $\Theta$ adjustable. For the Helmholtz coil geometry, $\Theta$ is easily adjustable by the current trough the coils. The easiest way to adjust $\Theta$ for a permanent magnetic setup is to change the distance between the neutron path (field box) and the magnets. To realize a homogeneous magnetic field density in the field box $\Omega_f$, the functional for the minimization problem can given by
\begin{align}
\label{eq:j_spin}
  J = \int_{\Omega_{f}} \left( | \triangledown B_x |^2 + | \triangledown B_y |^2	  + | (\triangledown B_z)_y |^2  + | (\triangledown B_z)_z |^2\right) \mathrm{d}  \boldsymbol{r}
\end{align}

\subsection{Two Candidate Designs}
\label{sec:spin_flipper_designs}

\begin{figure}[t]
	\centering
	\includegraphics[width=0.48\textwidth]{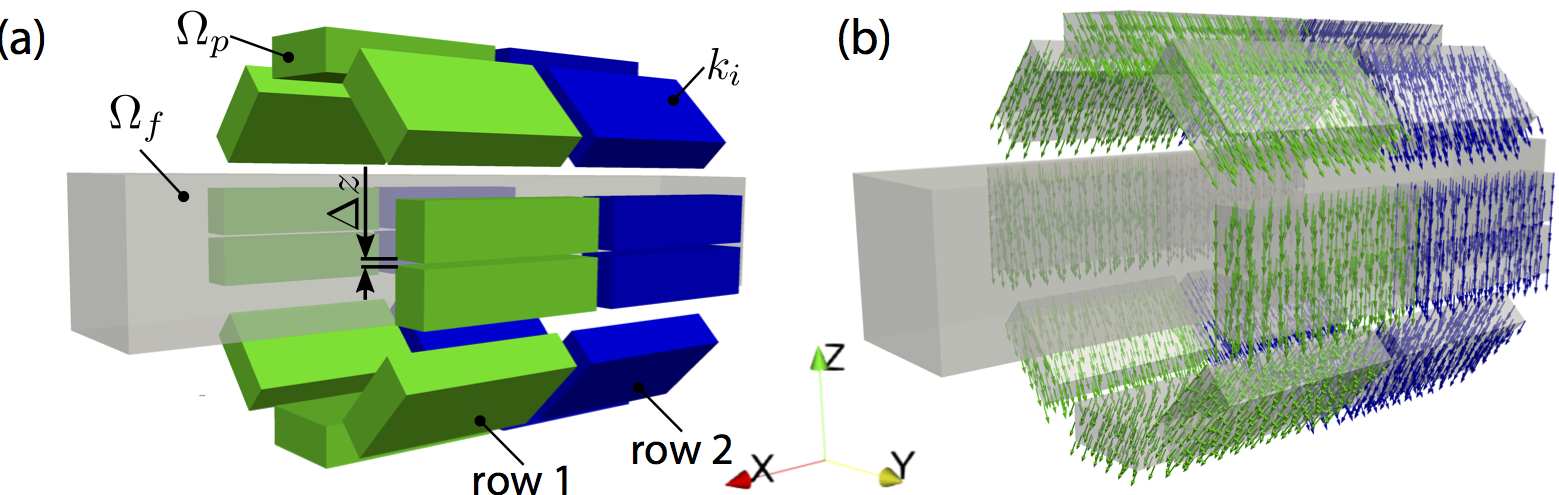}
	\caption[Modified and optimized Halbach cylinder.]{Modified and optimized Halbach cylinder. (a) Initial layout with 20 segments. Remanence $|\boldsymbol{B_r}|$ is constant for each segment. Two rows should provide a homogeneous field outside of the magnets. (b) Result of the inverse stray field computation. }
	\label{fig:spinflipper_new_design_result}
\end{figure}

\begin{figure}[b]
	\centering
	\includegraphics[width=0.48\textwidth]{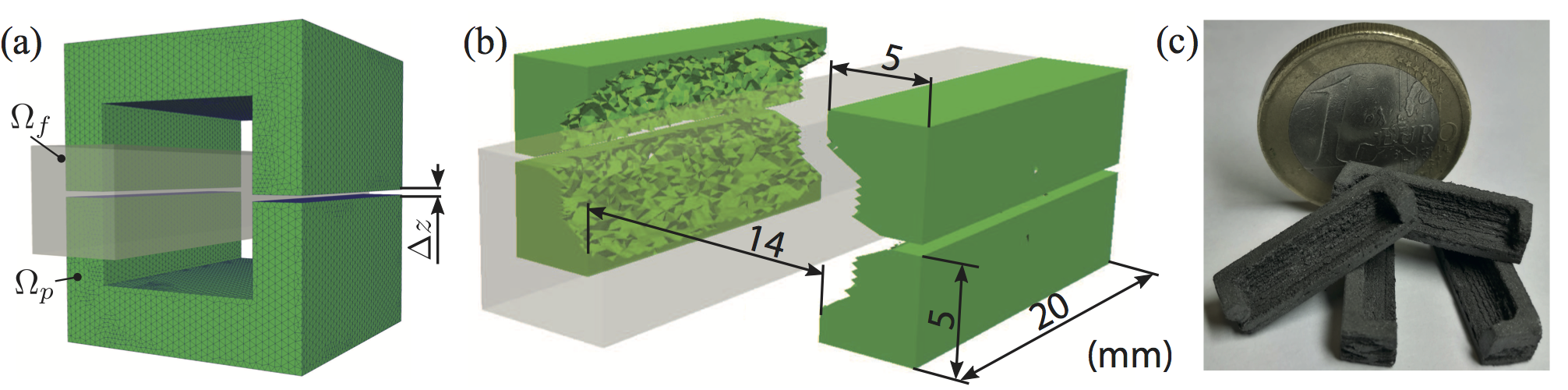}
	\caption[Topology optimized design for the Larmor spin-rotator.]{Topology optimized design for the Larmor spin-rotator. (a) Design domain for the topology optimization approach. The action $\Theta$ is adjustable by the gap size $\Delta z$. (b) Topology after optimization. (c) Photo of the magnets printed with Neofer$^\circledR$~25/60p.}
	\label{fig:spinflipper_new_design}
\end{figure}

Two different initial designs are investigated to find a proper replacement of the Helmholtz coils. The first design is a modified Halbach cylinder \cite{bjork2010comparison}. To achieve a better field homogeneity outside the magnet, it is divided into two rows, which is schematically illustated in Fig.\,\ref{fig:spinflipper_new_design_result}. To adjust $\Theta$, the design is split in the middle and the gap $\Delta z$ between both halves is adjustable. In total 20 segments are used. Each permanent magnetic segment has a constant remanence $|\boldsymbol{B_r}|$, but the direction of $\boldsymbol{B_r}$ is open and defines the optimization parameter for the inverse stray field optimization. Fig.\,\ref{fig:spinflipper_new_design}(a) shows the initial design of the modified Halbach cylinder. Due to the fact that only the direction of the remanence is an optimization parameter, no regularization parameter is necessary. Fig.\,\ref{fig:spinflipper_new_design_result}(b) shows the result of the inverse stray field simulation for an action of $\Theta=35$~mT$\cdot$mm. In general, the magnetization (remanence) vectors have the same direction as a standard Halbach cylinder, only the segments on the top and bottom of the field box show a deviating direction, to make the field in the field box more homogeneous.

The second investigated design is a topology optimized one. Fig.\,\ref{fig:spinflipper_new_design}(a) shows the design domain $\Omega_p$ and the field box $\Omega_f$ where $J$ of equation~\eqref{eq:j_spin} should be minimized. To adjust $\Theta$, the design consists of two halves, and the gap $\Delta z$ is adjustable. The mesh of the design domain consist of $256,542$ tetrahedral elements. No volume constraint is applied for the optimization. Fig.\,\ref{fig:spinflipper_new_design}(b) shows the topology optimized version of a permanent magnetic Larmor spin-rotator.

\begin{figure}[htbp]
	\centering
	\hspace{-2mm}\includegraphics[width=0.49\textwidth]{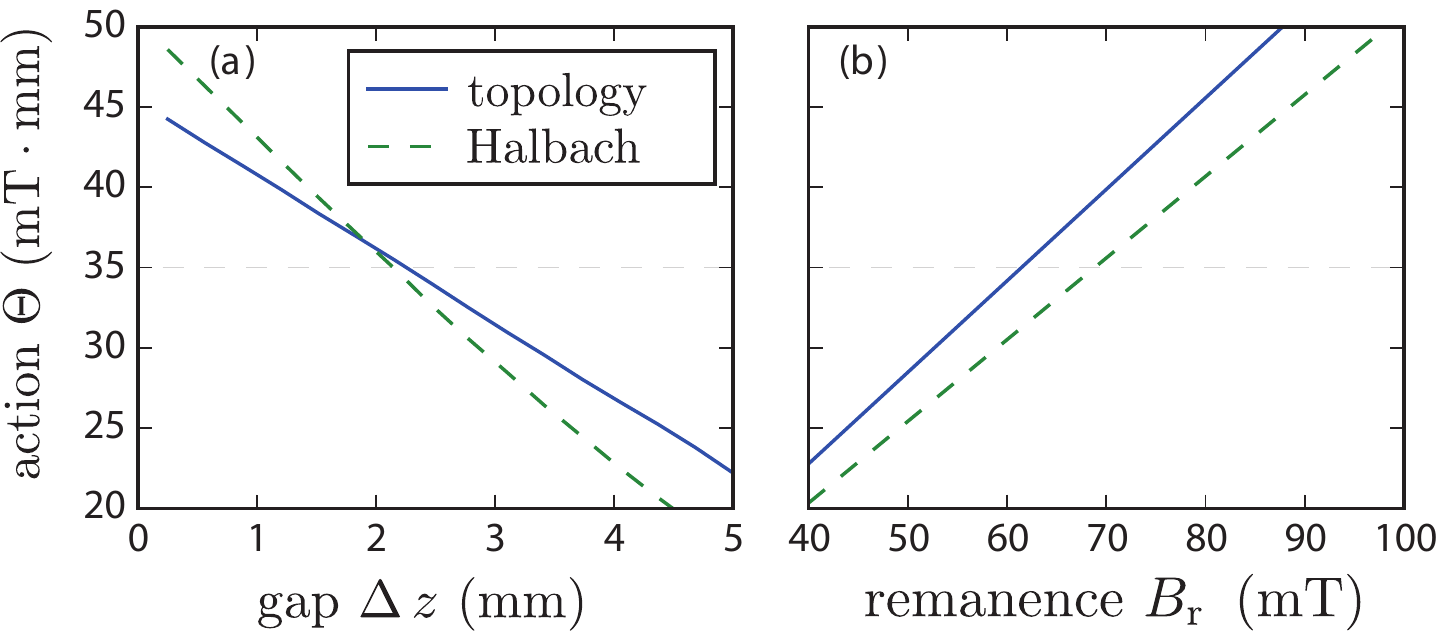}
	\caption[Action as a function of the gap and the remanence.]{Simulation results for both optimized designs. (a) Action $\Theta$ as a function of the gap $\Delta z$ ($B_r=61$~mT for the topology optimized design and  $B_r=68$~mT for the Halbach cylinder), and (b) as a function of the remanence $B_r$ of the magnet (gap $\Delta z=2$~mm).}
	\label{fig:gap_mag_action}
\end{figure}

Several numerical simulations are performed to find the optimal parameters of both designs. To adjust $\Theta$ in both directions, a gap of $\Delta z=2.25$~mm is chosen. To get an action of $\Theta=35$~mT$\cdot$mm, a remanence of the permanent magnet of $B_r=61$~mT for the topology optimized version and $B_r=68$~mT for the Halbach design is necessary. How $\Delta z$ and $B_r$ influence $\Theta$, is plotted in Fig.\,\ref{fig:gap_mag_action}. There exists a linear correlation between action and gap size. The topology optimized design is less sensitive for changing $\Delta z$.

\begin{figure}[!h]
	\centering
	\hspace{-2mm}\includegraphics[width=0.49\textwidth]{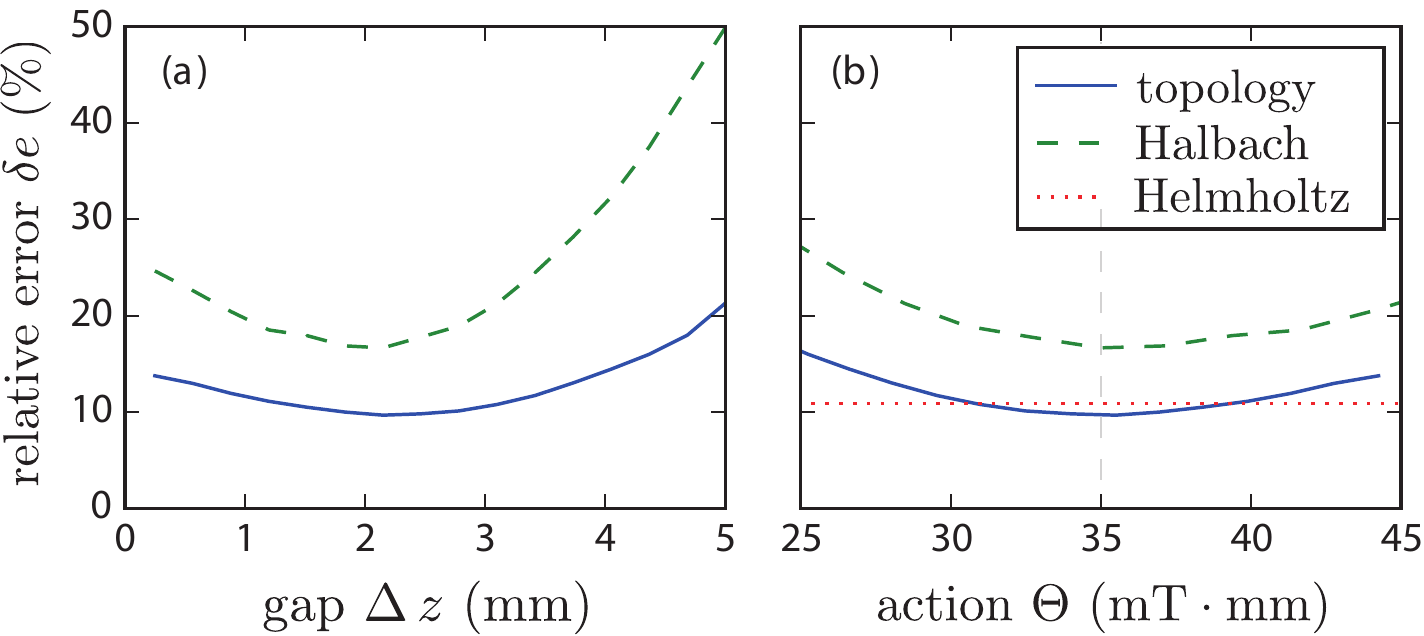}
	\caption[Relative error for both optimized designs.]{Relative error for both optimized designs. (a) Relative error $\delta e$ as a function of the gap $\Delta z$. (b) Relative error as a function of the action $\Theta$ for the optimized designs and for the current Helmholtz coil geometry.
	}
	\label{fig:gap_action_relerror}
\end{figure}

The homogeneity of $\boldsymbol{B}(\boldsymbol{r})$ in the field box $\Omega_f$ has a crucial impact on the performance. The relative error is defined as
\begin{align}
 \delta e = \frac{J}{ \int_{\Omega_f} |\boldsymbol{B}|^2 \mathrm{d}  \boldsymbol{r} } 
\end{align}
with the functional $J$ of equation~\eqref{eq:j_spin}. Fig.\,\ref{fig:gap_action_relerror}(a) shows the relative error $\delta e$ as a function of the gap $\Delta z$. The topology optimized design has a much lower relative error compared to the Halbach cylinder design, and the dependency on the gap size is much lower as well. However, the main question is, if the optimized permanent magnetic design has a lower relative error, or a better performance as the current Helmholtz coil geometry. Fig.\,\ref{fig:gap_action_relerror}(b) shows a plot of the relative error as a function of the action for the current and the both optimized designs. For the current design, $\delta e$ is independent of $\Theta$. In the range of around of $\Theta=30-40$~mT$\cdot$mm the topology optimized version shows a better performance. This translates to a range of $\Delta z \simeq 1 - 3.5 ~\text{mm}$ for the tunable gap distance.

\begin{figure}[htbp]
	\centering
	\includegraphics[width=0.45\textwidth]{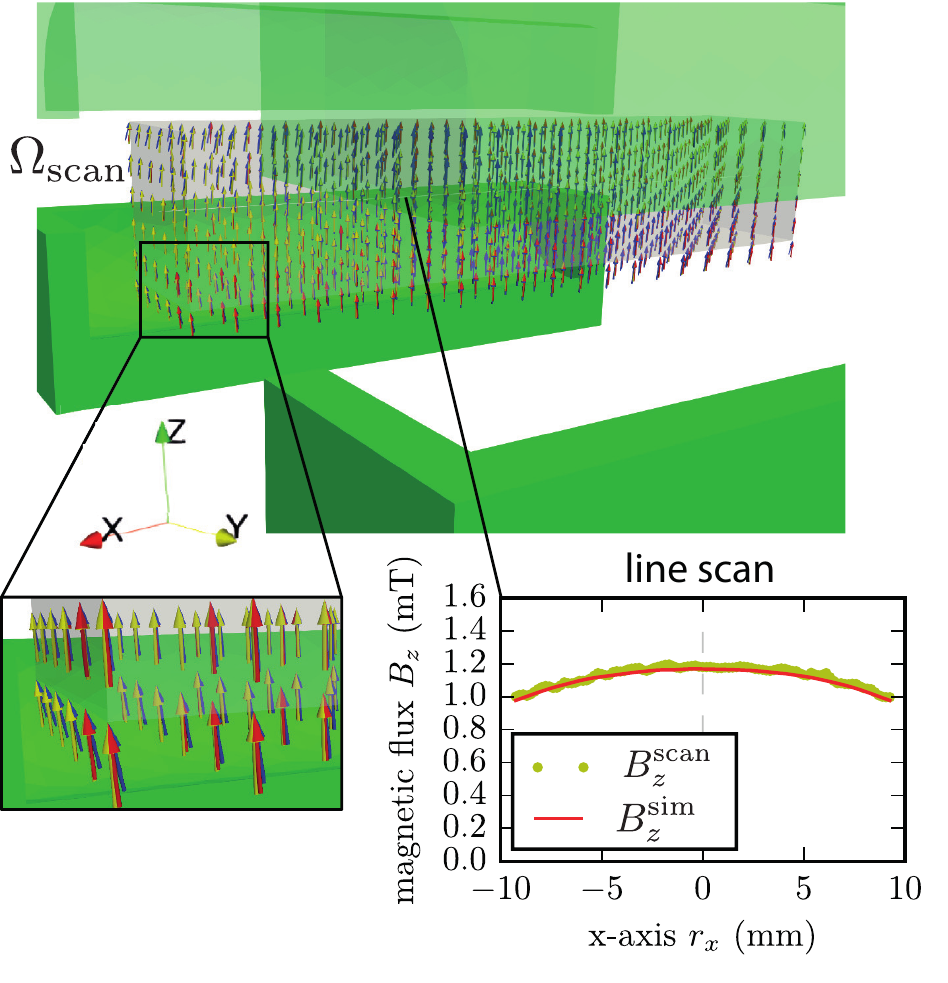}
	\caption[3D printed Larmor spin-rotator and test of the magnetic performance.]{Simulation and measurement of the magnetic field. Volume scan between the magnets for a gap of $\Delta z= 2.25$~mm. Magnetic vectors (blue: coil simulation, red: topology optimization simulation, yellow: measurement) illustrate the homogeneity of the field. Line scan of $B_z$ at $y=0$~mm, $z=0$~mm.}
	\label{fig:spin_flipper_2}
\end{figure}
\begin{figure*}[htbp]
	\centering
	\includegraphics[width=1\textwidth]{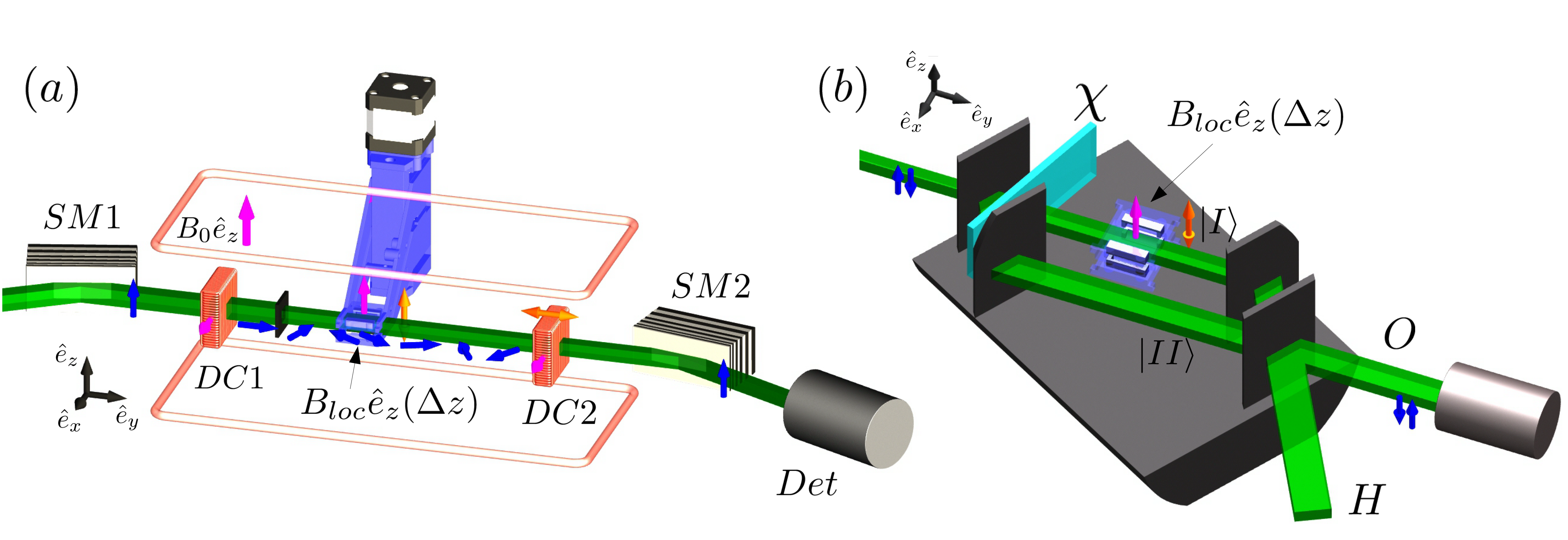}
	\caption[Testsetup to adjust Larmor precession angle of 3D printed Larmor spin-rotator neutron interference experiments.]{3D printed  Larmor spin-rotator neutron interference experiments. (a) Illustration of the polarimetric setup. Polarized neutrons are flipped into (out of) the x-y-plane by direct current spin-rotators DC1 (DC2), where they rotate in the guide field $B_0$ due to Larmor precession. Arrows indicating spin directions (blue) and magnetic fields (magenta). The 3D printed magnets create an additional z-field $B_{loc}(\Delta z)$ that can be controlled by a motorized adjustment mechanism. The position of DC2 can be varied along the beam. After the spin direction is analyzed by the supermirror SM2, the neutrons are detected in a He-3 counter tube (Det). (b) Illustration of the interferometric setup. The monochromatized, unpolarized neutron beam is split in two parts, path $\ket{I}$ and $\ket{II}$. Their relative phase can be adjusted by rotating the phase shifter plate ($\chi$). The 3D printed magnets with variable $\Delta z$ are placed in path $I$. At the last interferometer plate, the two paths recombine and neutrons in the O beam pass on into the detector.
	}
	\label{fig:exp_setups}
\end{figure*}

\subsection{Fabrication}
\label{sec:spin_flipper_validation}
For the manufacturing process of the permanent magnets, we use a conventional low-cost end-user 3D printer without any modifications \cite{huber20163d} to print the topology optimized system. As a printing material the prefabricated compound material (Neofer$^\circledR$~25/60p) from Magnetfabrik Bonn GmbH is used to realize the setup. All four segments have the same shape. Fig.\,\ref{fig:spinflipper_new_design}(c) shows a picture of the printed segments. After the printing process, the magnets must be magnetized. Compared to all previous applications, only a weak magnetic field is necessary. The segments must have exactly a remanence of $B_r=61$~mT to generate an action of $\Theta=35$~mT$\cdot$mm for a gap of $\Delta z=2.25$~mm. 
Magnetization of the segments are performed with an electromagnet with a maximum magnetic flux density inside the electromagnet of $1.9$~T in permanent operation mode. A jig with the exact positions of the segments is 3D printed. The jig is inserted into the electromagnet and the external field is increased in small steps. After each step, the magnetic field density  of the segments is measured by the 3D Hall probe. FEM simulation of the  arrangement yields a field of $B_z=1.18$~mT in the center for the desired action $\Theta$. With this approach a good adjustment of the remanence $B_r$ is possible.

After the magnetization procedure, the magnetic field between the segments is measured and compared with simulation results. Fig.\,\ref{fig:spin_flipper_2}(b) shows a volume scan between the segments and a line scan  of $B_z$ at $y=0$~mm, $z=0$~mm for the measured topology optimized version and the former Helmholtz coil geometry. Simulation results for $B_r=61$~mT and measurements are in a good agreement. The vector field illustrates the homogeneity of the measurement and simulations, respectively.

\begin{figure}[b]
	\centering
	\includegraphics[width=0.5\textwidth]{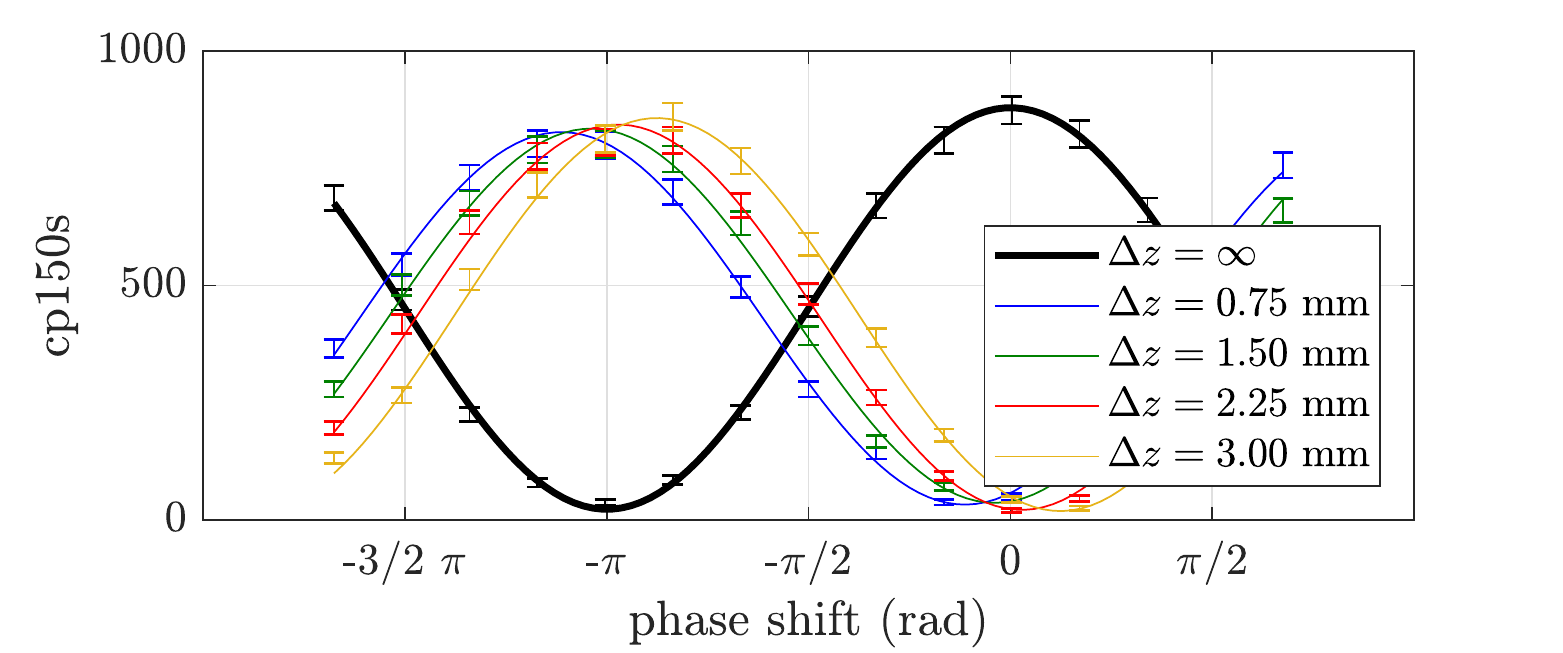}
	\caption[Polarimetric phase shifts]{
		Polarimetric test measurements of the spin-rotation angle caused by the 3D printed magnets with remanence $B_r = 61$ mT. Interference patterns for different $\Delta z$ attained by translating DC2 along the beam.
	}
	\label{fig:spin_flipper_result}
\end{figure}

\section{Performance in neutron optical experiments}
\subsection{Polarimeter experiment}
The real performance of the 3D printed Larmor spin-rotator can be only tested with a neutron experiment. For this reason, interference measurements with neutrons are performed at the TRIGA MARK-II reactor at the Atominstitut. Fig.~\ref{fig:exp_setups} shows an illustration of the experimental setups. The gap $\Delta z$ and therefore the action $\Theta$ can be adjusted by a 3D printed mounting system with counter-rotating threads. In the polarimeter experiment, depicted in Fig.\,\ref{fig:exp_setups}(a), the first direct current spin turner DC1 rotates the spin of the neutrons in flight direction by an angle of $\pi/2$, then a cadmium aperture reduces the beam. After the aperture, the beam enters the 3D printed Larmor spin-rotator. The intensity modulation is created by varying the position of DC2, which causes another $\pi/2$ spin-rotation around the x-axis. After DC2, the neutrons pass through a supermirror analyzer, transmitting only the $\ket{+z}$-spin component into the detector.

Interference patterns for different $\Delta z$ are plotted in Fig.\,\ref{fig:spin_flipper_result}(b). These are only test measurements with a small aperture opening of $3\times5$~mm$^2$. The reference measurement (black curve), from which the phase shifts are measured, is done without the Larmor spin-rotator. These measured phase shifts directly translate to the spin-rotation angle $\alpha$. First measurements show a spin-rotation of $\alpha=\pi$ for a gap of $\Delta z=2.25$~mm with the original remanence of $B_r = 61$ mT. This fits really well to the simulation for this action $\Theta$. Another crucial parameter is the spin contrast $C_S$ of the setup, which can be calculated from $C_S = \frac{I_0 - I_\pi}{I_0 + I_\pi}$, where $I_\pi$ ($I_0$) is the intensity behind the spin analyzer after a $\pi$-flip (no flip). Here, a contrast of more than $95$~\% can be achieved for a gap of $\Delta z=2.25$~mm.

\begin{figure}[!b]
	\centering
	\includegraphics[width=0.5\textwidth]{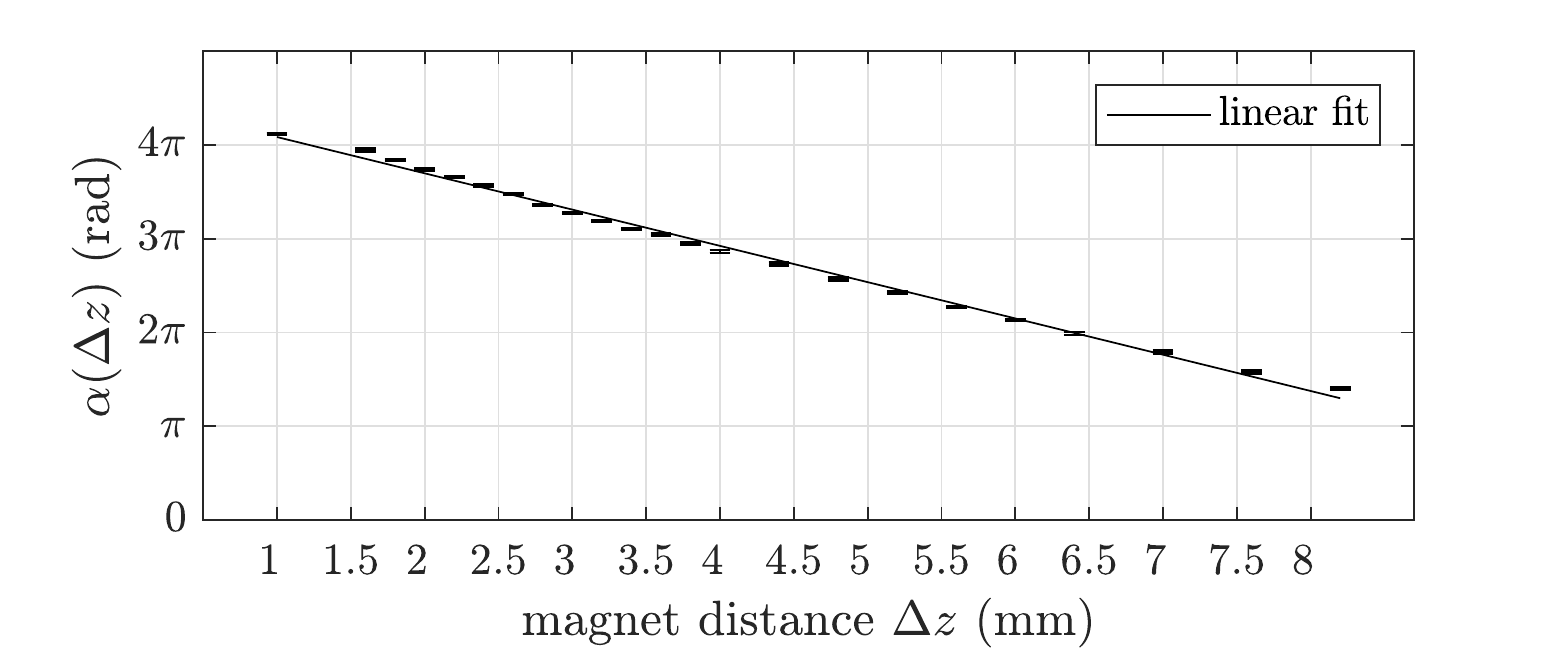}
	\caption[Polarimetric phase shifts]{
		Later polarimetric measurements with four times higher remanence $B_r = 244$ mT. Spin-rotation angle $\alpha$ of the measured interference patterns calculated from sine fits against magnet distance $\Delta z$. The reference measurement is $\Delta z = \infty$, i.e. magnets taken out of the beam.
	}
	\label{fig:polarimeter_result}
\end{figure}

However, the measurements showed such a promising linearity of the phase shift with $\Delta z$, that the 3D printed magnets were magnetized to a four times higher remanence of $B_r = 244$ mT, in order to have a wider tuning range of the spin-rotation angle $\alpha$. Fig.\,\ref{fig:polarimeter_result} shows plots of $\alpha$ calculated from phase shifts of the interference patterns against $\Delta z$. These polarimetric measurements are carried out with an improved motorized adjustment mechanism for the gap distance and a wider aperture of $7\times 7$~mm$^2$. They show a good linearity of the spin-rotation angle against the gap size.


\begin{figure}[!t]
	\centering
	\includegraphics[width=0.5\textwidth]{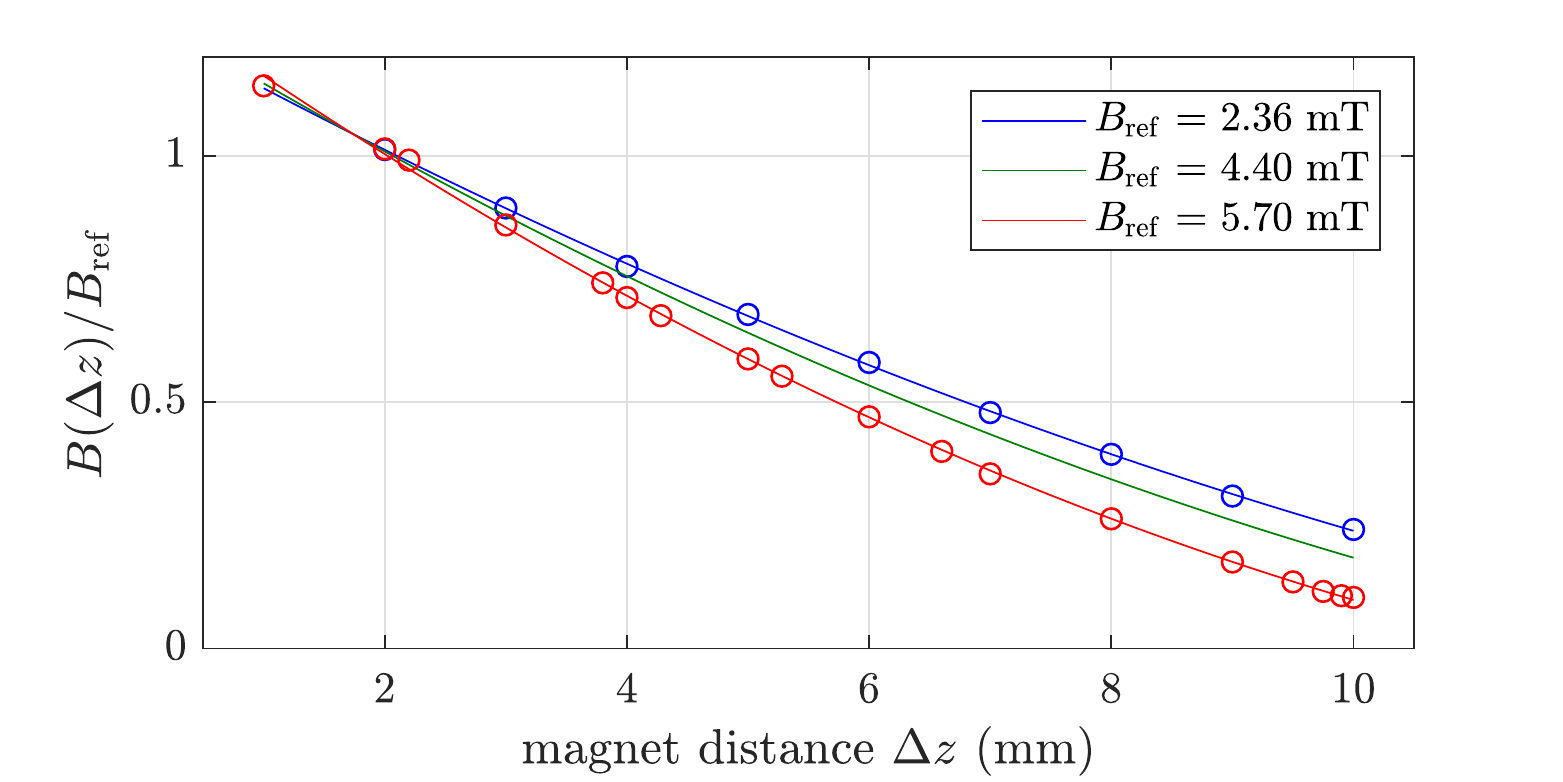}
	\caption[Polarimetric phase shifts]{
		Plot of the magnetic field at the center of the 3D printed Larmor-spin rotator, normalized by the value $B_{\rm{ref}}$ measured at $\Delta z = 2.25$ mm. For $B_{\rm{ref}} = 2.36$ mT and $B_{\rm{ref}} = 5.70$ mT the magnetic field has been measured using a Hall probe and fitted with a quadratic polynomial. The curve for $B_{\rm{ref}} = 4.40$ mT is interpolated using these fits.
	}
	\label{fig:result_magnetfeld}
\end{figure}

\subsection{Interferometer experiment}
As an example of the performance of the newly fabricated spin-rotators at their intended application inside a perfect crystal neutron interferometer, measurements are carried out at the Atominstitut. The experiment is comparable to the first demonstration of the $4\pi$-spinor symmetry of neutrons \cite{rauch1975fourpi}, its setup is depicted in Fig.\,\ref{fig:exp_setups}(b). The unpolarized neutron beam with a cross section of $10 \times 10$~mm$^2$ is split in path $\ket{I}$ and $\ket{II}$ after the first interferometer plate. The phase shifter plate can be rotated in order to create a phase difference $\chi$ between the two paths. The 3D printed Larmor spin-rotator with variable $\Delta z$ is placed in path $I$. At the last interferometer plate, the two paths are recombined and leave the interferometer in two separate beams, the O beam in the forward direction and the reflected H beam. Only the O beam has the same number of reflections and transmissions, and therefore is able to show maximum interferometric contrast $C$, when the phase shift $\chi$ is varied. The O beam intensity of an empty interferometer as a function of $\chi$ can be written as \cite{RauchBook}
\begin{align}
	I_O(\chi) = \lvert \Psi_0 \rvert^2 (1 + C \cos\chi).
\end{align}
Here, $C$ is an parameter and depends on the interferometer used, temperature gradients, vibrations, etc. The density matrix of an unpolarized neutron in the interferometer, assuming perfect contrast $C=1$, can be written as the direct product
\begin{align}
	\rho = \frac{1}{2}\dyad{\psi_i}\otimes\frac{1}{2}(\ket{\uparrow}\!\!\bra{\uparrow}+\dyad{\downarrow}),
\end{align}
where the first term in the product is the path state $\ket{\psi_i} = \frac{1}{\sqrt{2}}(\ket{I} + e^{i \chi}\ket{II})$, which depends on the phase shift $\chi$, and the second term is the maximally mixed spin state. The interaction with the local magnetic field in path $I$ can be modeled as
\begin{align}
	U_{int}(\alpha) = \dyad{I}\otimes e^{i \frac{\alpha}{2} \sigma_z} + \dyad{II} \otimes \mathbbm{1},
\end{align}
where $\alpha$ depends on $B_{loc}(\Delta z)$ and is given by equation \eqref{eq:alpha}, when B is homogeneous. At the last interferometer plate, the projector $P_{f}=\dyad{\psi_f}\otimes\mathbbm{1}$ acts on the path state and projects it onto the state $\ket{\psi_f} = \frac{1}{\sqrt{2}}(\ket{I} + \ket{II})$. Therefore, the intensity at the O detector is given by
\begin{equation}
\begin{split}
I_O(\chi,\alpha(\Delta z)) 
 &\propto \frac{1}{C}\Tr(P_{f} U_{int} \rho U_{int}^\dagger P_{f}^\dagger) + \frac{C-1}{C} \\
&= \frac{1}{2}\left(1 + C \cos(\frac{\alpha}{2})\cos(\chi)\right),
\end{split}
\end{equation}
where the empirical prefactor $C$ has been reintroduced in front of the interference term. The interferometric contrast $C\cos(\alpha/2)$ is reduced to zero for odd multiples of the spin rotation angle $\alpha = \pi$, i.e., when the spin states in the two paths are orthogonal. At $\alpha = 2 \pi$ the interferogram exhibits a phase shift of $\pi$ compared to $\alpha = 0$, i.e. magnets removed. Only at $\alpha = 4\pi$ the interferogram has returned to its initial contrast and phase, which shows the $4\pi$-symmetry of the neutron as a spin-$1/2$ particle. 

\begin{figure}[htbp]
	\centering
	\includegraphics[width=0.5\textwidth]{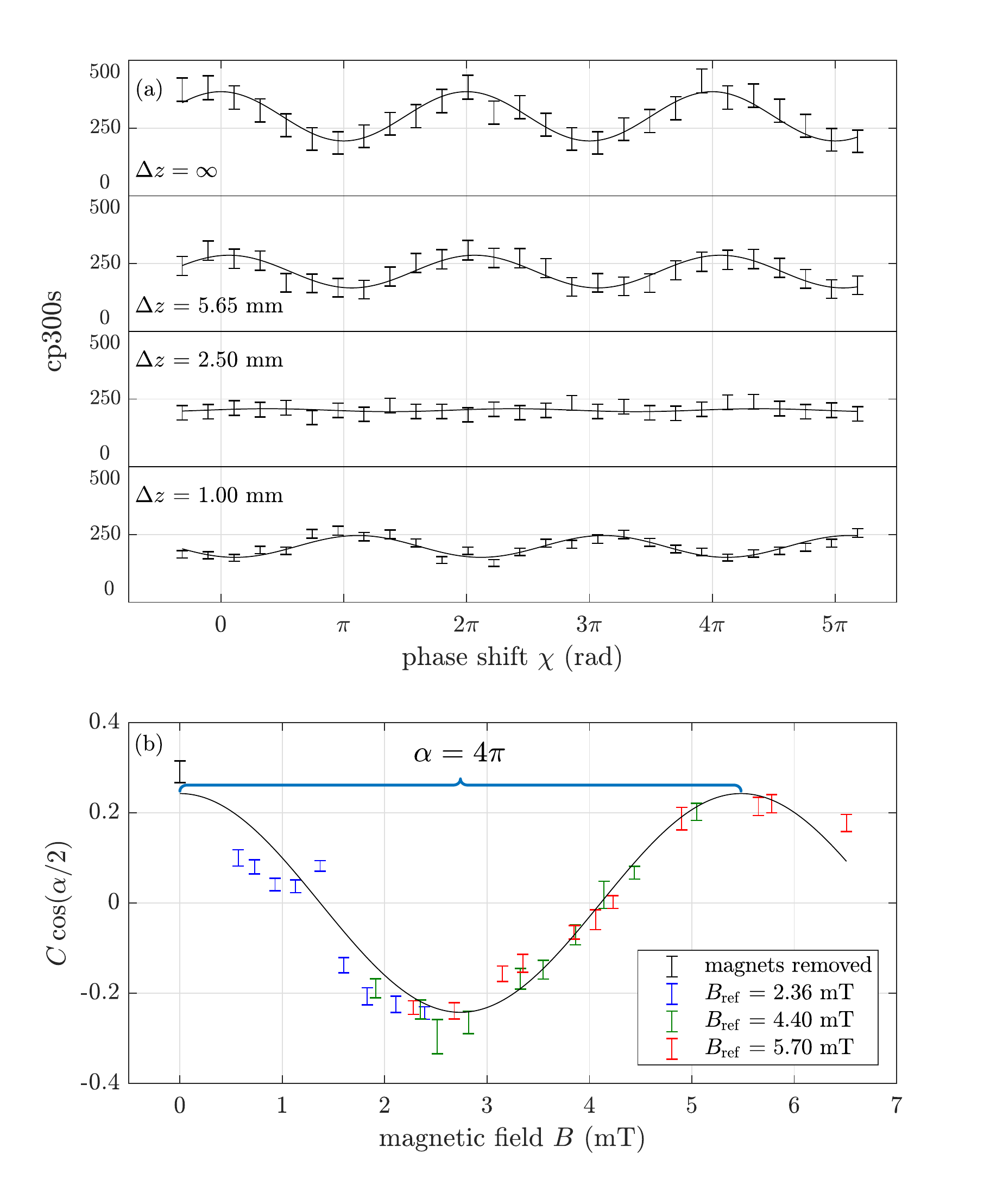}
	\caption[Polarimetric phase shifts]{
		Interferometric measurement results. (a) Typical interferograms for different values of $\Delta z$. The curves shown have been measured for $B_{\rm{ref}}$ = 4.40 mT. (b) Plot of $C \cos(\alpha/2)$ extracted from fits to the recorded interferograms against the magnetic field at the center of the 3D printed Larmor spin-rotator. Different data points have been measured using different magnetization strengths.
	}
	\label{fig:interferometer_result}
\end{figure}

In order to experimentally demonstrate this $4\pi$-symmetry in a neutron interferometer experiment, three different magnetization strengths of the 3D printed Larmor spin-rotator are used, each identified by the magnetic field strength $B_{\rm{ref}}$ measured in the center point at $\Delta z = 2.25$ mm. Measurements of the dependence of the magnetic field on the distance $\Delta z$ are plotted in Fig.\,\ref{fig:result_magnetfeld}. For distances upwards of 8 mm, the magnetic field ceases to be linear. Nevertheless, interferograms are recorded for $\Delta z$ up to 10 mm and plotted against the magnetic field calculated from fits with a quadratic polynomial. In Fig.\,\ref{fig:interferometer_result} the results of the interferometric measurements are depicted, which clearly reproduce the $4\pi$-symmetry in the experiment.

\section{Discussion and Outlook}
The combination of 3D printing and topology optimization using FEM methods offers new possibilities for tackling design problems, e.g. restrictions in available space to implement physical interactions. The design goal of this work, a Larmor spin-rotator consisting of a region of homogeneous magnetic field density, which is also tunable in magnitude, is met by a setup of four segments which are located to the sides of the field box. Tunability of the magnetic field strength is realized by introducing a variable gap between the top and bottom half of the spin-rotator. The new method can be used in applications, where electromagnetic coils are unfavorable, e.g., due to heat dissipation or undesired inductances. Future developments could include different field configurations. One example is a spin-rotation field, which is perpendicular to the guide field and a non-adiabatic field transition is required. Another possible field configuration is that of a wiggler, which consists of stacked regions of anti-parallel magnetic fields along the beam-path. The spatial variation of the magnetic fields in the wiggler leads to resonant Rabi-flops of the spin, similar to the working principle of a resonant-frequency spin-flipper, where the variation happens in the time-domain. Using such devices, it is also possible to shape the beam-profile in momentum space. The new Larmor spin-rotator design developed in this work can be seen as a first step toward an implementation of custom magnetic field shaping using 3D printed permanent magnets in neutron optics.

\section{Conclusion}
We have developed a new method for coherent Larmor spin-rotation control using topology optimized 3D printed magnets for applications in neutron interferometer experiments, where the rotation axis is parallel to the outer guide field. The magnetic action can be adjusted linearly by varying the distance between the magnets. We have showed that spin-rotation angles of more than $4 \pi$ are possible, depending on the initial magnetization strength. Measurements are in good agreement with simulations and showed a spin contrast of more than 95\%, comparable to older methods using Helmholtz coils. The advantages of this new method are that unwanted inductances are avoided, and that no heat is dissipated by current carrying wires, which prevents a reduction in interferometric contrast due to temperature gradients or would make water cooling necessary. 


%

\end{document}